\documentstyle[twocolumn,prl,epsf,aps]{revtex}

\newcommand{\rmd}{{\rm d}}
\newcommand{\rmi}{{\rm i}}

\newcommand{\beq}{\begin{equation}}
\newcommand{\eeq}{\end{equation}}
\newcommand{\bea}{\begin{eqnarray}}
\newcommand{\eea}{\end{eqnarray}}

\newcommand{\cuoo}{CuO$_{2}$}
\newcommand{\oco}{O-Cu-O}
\newcommand{\lsco}{La$_{2-x}$Sr$_{x}$CuO$_{4}$}
\newcommand{\ybco}{YBa$_{2}$Cu$_{3}$O$_{6+x}$}

\begin{document}                                                

\wideabs{

\draft

\title{Charge pairing and superconductivity in high-T$_{c}$ cuprate superconductors}

\author{Eduardo C. Marino and Marcello B. Silva Neto}

\bigskip

\address{
Instituto de F\protect\'\i sica, Universidade Federal do Rio de Janeiro,
Caixa Postal 68528, Rio de Janeiro - RJ, 21945-970, Brazil}

\date{\today}
\maketitle


\begin{abstract} 

We propose a model for high-T$_{c}$ superconductors 
that includes both the spin fluctuations 
of the Cu$^{++}$ magnetic ions and of the spins of 
O$^{--}$ doped holes (spinons). The charge of the 
dopants (holons) is associated to quantum skyrmion 
excitations of the Cu$^{++}$ spin background. 
The quantum skyrmion effective interaction 
potential is evaluated as a function of doping and
temperature, indicating that Cooper pair 
formation is determined by the competition 
between these two types of spin fluctuations. The 
superconducting transition occurs when the effective 
potential allows for skyrmion bound states. 
Our theoretical predictions for the superconducting 
phase diagram of {\lsco} and {\ybco} are in good 
agreement with experiment. 

\end{abstract}


}               

\begin{narrowtext}


{\it Introduction.}
High-temperature underdoped cuprates exhibit 
a wide variety of interesting physical phenomena, 
like N\'eel and metal-insulator transitions, non-Fermi 
liquid behavior, pseudogap, etc., and have 
inspired a large amount of theoretical and 
experimental work for about fifteen years. 
In spite of that, even the nature of the 
ground state and of its elementary excitations, 
have not yet been fully determined and many 
different pictures are availabe \cite{Review}. 

Another fundamental point yet to be understood 
is the mechanism of charge pairing. It is by now 
well established that antiferromagnetic spin 
correlations play an important role in the 
dynamics of the system, even after the destruction 
of the N\'eel state. Indeed, different spin-fluctuation 
models have been successfully used to explain the 
observed spectral weight in ARPES data of high-T$_{c}$ 
materials \cite{Spectral-Weight}, as well as other 
anomalies \cite{Anomalies}. Moreover, the idea of 
spin-fluctuation induced charge pairing and 
superconductivity has been used recurrently 
\cite{Spin-Fluctuation}. 

In this work we propose a theory for high-T$_{c}$ 
cuprates that takes into account 
the spin fluctuations of the Cu$^{++}$ 
magnetic ions and of the O$^{--}$ doped 
holes as independent degrees of freedom. The 
charge of the dopants (holons) is associated 
to skyrmion quantum spin excitations of 
the Cu$^{++}$ background, which in the 
N\'eel phase are finite energy defects 
closely related to their classic counterparts 
whereas in the quantum disordered phase are 
nontrivial zero energy purely quantum mechanical 
excitations. The spin of the doped holes (spinons), 
on the other hand, is represented 
by chargeless, massless Dirac fermion fields
\cite{Marston-Affleck}. We calculate 
the effective interaction potential between 
these quantum skyrmion topological 
excitations in order to study charge 
pairing. It becomes clear that 
Cooper pairing is controlled by 
the competition between the spin 
fluctuations of Cu$^{++}$ magnetic ions and those of the 
O$^{--}$ doped holes. Our predictions for the
$T_{SC}$ line are in good agreement with experiment
for both {\lsco} and {\ybco}.

{\it The model.}
Our starting point will be the generalized 
spin-fermion model described by the
Hamiltonian
\bea
{\cal H}&=&-t_{p}\sum_{\left<i,j\right>,\alpha}
(c^{\dag}_{i,\alpha}c_{j,\alpha}+h.c.)+
U_{p}\sum_{i,\alpha}n_{i,\alpha}n_{i,\alpha} \nonumber \\
&+&J_{K}\sum_{i,\alpha,\beta}\vec{S}_{i}
\cdot c^{\dag}_{i,\alpha}{\vec \sigma}_{\alpha\beta}c_{i,\beta}+
J\sum_{\left<i,j\right>}\vec{S}_{i}\cdot\vec{S}_{j},
\label{SF-Hamiltonian}
\eea
which arises from the strong coupling limit 
of the three band Hubbard Model ($3$BHM) 
\cite{Emery}. In the above expression, $\vec{S}_{i}$ 
represent the localized spins of Copper ions, which 
interact through the superexchange 
$J$, $c^{\dag}_{i,\alpha}$, $\alpha=1..N=2$, 
is the hole creation operator, $t_{p}$ is the hopping 
term for holes, $J_{K}$ is a Kondo like coupling 
between the spins of Cu$^{++}$ ions and the spins of
O$^{--}$ holes, and we have retained the usually 
ignored onsite Coulomb repulsion between O$^{--}$ holes, 
$U_{p}\neq 0$ with $n_{i,\alpha}=c^{\dag}_{i,\alpha}c_{i,\alpha}$. 
The reason is that realistic estimates from the 
$3$BHM suggest that $U_{p}/t_{p}\sim 10$ \cite{Kampf}, 
being rather large, and thus we can perform a $t_{p}/U_{p}$ 
expansion. Second order perturbation theory
in $t_{p}/U_{p}$ will give rise to a superexchange 
$J_{p}=2t_{p}^2/U_{p}$ between oxygen spins and we end 
up with a $t-J$ model for the holes.

The mean field (large $N$) solutions of 
the $t-J$ model are well known and it has been 
established that a $\pi$-flux phase has 
minimum energy, at least at the saddle-point level 
($N\rightarrow\infty$) \cite{Marston-Affleck}. 
We can write the electron in terms of a 
charged spinless boson $\mu_{i}$ (holon) and 
a chargeless spin-$1/2$ fermion $f_{i,\alpha}$
(spinon), and, as usual, we decouple the
four particle interactions by introducing the
$d$-wave auxiliary fields 
$\chi_{ij}=\left<f^{\dag}_{i,\alpha}f_{j,\alpha}\right>$ and
$\Delta_{ij}=\left<f_{i,\uparrow}f_{j,\downarrow}-
f_{i,\downarrow}f_{j,\uparrow}\right>$, which
are nonzero for $T<T^{*}$, where $T^{*}$ is the
pseudogap temperature. If we then neglect charge 
fluctuations, $\left<\mu_{i}\mu^{\dag}_{j}\right>\simeq 
\left|\mu_{i}\right|^{2}=const.$, we find that the lowest 
lying excitations of the $\pi$-flux phase 
are massless, chargeless, spin 
carrying Dirac Fermi fields \cite{Marston-Affleck} 
whose dynamics is described by the Lagrangian 
${\cal L}=\sum_{\alpha,\lambda}
\rmi\overline{\psi}_{\alpha,\lambda}
\left(\gamma_{0}\partial_{\tau}-
v_{F}\vec{\gamma}\cdot\vec{\nabla}
\right)\psi_{\alpha,\lambda}$, where $\lambda=1,2$ label
the two Fermi points at $(\pi/2,\pm\pi/2)$,
$\partial_{\mu}=(\partial_{\tau},\vec{\nabla})$, 
$\gamma_{\mu}=(\gamma_{0},\vec{\gamma})=
(\rmi\sigma_{z},\sigma_{x},\sigma_{y})$, 
$v_{F}=2 a \chi$ is the dopant Fermi velocity (with
$a$ being the lattice spacing and $\chi$ the 
constant amplitude of $\left|\chi_{ij}\right|$)
and $\psi_{\alpha,\lambda}=\pmatrix{f^{e}_{\alpha,\lambda}
\cr f^{o}_{\alpha,\lambda}}$, for ($o$)dd and ($e$)ven 
lattice sites. The long wavelength fluctuations of
the localized Cu$^{++}$ spins, on the other hand, are
described by the CP$^{N-1}$ Lagrangian
${\cal L}_{CP^{N-1}}=(1/2g_{0})
\left|(\partial_{\mu}-\rmi{\cal A}_{\mu})z_i\right|^{2}$,
where $\vec{S}=z_{i}^{\dag}${\mathversion{bold}
$\vec{\sigma}$}$_{ij} z_{j}$, with $z_i^{\dag},z_i$, 
$i=1..N=2$, being Schwinger boson fields such
that $z^{\dag}_{i}z_{i}=1$, ${\cal A}_{\mu}=-
\rmi\bar{z}_{i}\partial_{\mu}z_{i}$, and $g_{0}$ 
is a bare coupling constant. It is now convenient 
to perform the local canonical transformation 
$\psi\rightarrow U\psi$, where 
$U=\exp{\left[q\pmatrix{z_{1}&-\bar{z}_{2}\cr 
z_{2}&\bar{z}_{1}}\right]}$ $\in$ $SU(2)$, and 
$q$ is arbitrary. Now the Kondo coupling term 
in (\ref{SF-Hamiltonian}) reduces to a chemical 
potential term, since $U^{\dag}\vec{S}\cdot\vec{\sigma} 
U=\sigma_{z}$. Also, since $U^{\dag}\partial_{\mu}U=
\rmi q\sigma_{z}{\cal A}_{\mu}+$ negligible nondiagonal 
terms, 
we end up with the effective 
theory
\beq
{\cal Z}=\int{\cal D}\bar{z}{\cal D}z{\cal D}
\overline{\psi}{\cal D}\psi
{\cal D}{\cal A}_{\mu}{\;}\delta[\bar{z}z-1]
e^{-S},
\label{Part-Func}
\eeq
where 
\bea
S&=&\int_{0}^{\beta\hbar}\rmd\tau
\int\rmd^{2}{\bf x}
\left\{\sum_{i=1..N}\frac{1}{2 g_{0}} 
\left|(\partial_{\mu}-\rmi{\cal A}_{\mu})z_i\right|^{2}
\right. \nonumber \\ &+& \left. 
\sum_{\alpha=1..N,\lambda=1,2}\overline{\psi}_{\alpha,\lambda} {\;} 
\gamma_{\mu} \left(\rmi\partial^\mu - q\sigma_{z}
{\cal A}^{\mu}\right)\psi_{\alpha,\lambda} \right\}.
\label{Effective-Action}
\eea

{\it Holons as quantum skyrmions.}
In previous works \cite{Marino-Marcello}, 
we have proposed a model for 
doping quantum Heisenberg antiferromagnets,  
that successfully described the magnetization 
curves and the AF part of the phase diagrams 
of both LSCO and YBCO. One of the 
important consequences of that model was the
observation that each hole added to the 
{\cuoo} planes creates a skyrmion topological 
defect on the Cu$^{++}$ spin background, 
in agreement with earlier proposals
\cite{Skyrmion-Literature}. The dopant 
charge, in particular, was found to be 
attached to the skyrmion charge and 
consequently its dynamics becomes totally 
determined by the quantum skyrmion 
correlation functions. Despite the fact that
the model proposed in \cite{Marino-Marcello} 
is restricted to the antiferromagnetic 
part of the phase diagram, we shall 
nevertheless pursue the picture in which 
skyrmions are in general the charge 
carriers of the doped holes. This will allow 
us to treat the bosonic variable 
$\mu_{i}$ introduced above as a quantum skyrmion 
operator. In particular, we shall exploit this 
idea in the quantum disordered phase, 
$\delta\geq\delta_{AF}$, where the skyrmions 
are purely quantum mechanical and have zero 
energy. 

The full treatment of the quantum 
skyrmions of the theory described by 
(\ref{Effective-Action}) has been 
carried out in \cite{Marino}. In the 
renormalized classical regime, 
$g_{0}<g_{c}$ ($g_{c}=8\pi/\Lambda$), we have
\beq
\langle\mu(x)\mu^{\dag}(y)\rangle=
\frac{e^{-2\pi\rho_{s}|x-y|}}{|x-y|^{q^2/2}},
\label{sk-cf1}
\eeq
where $\rho_{s}=1/g_{0}-1/g_{c}$.
Conversely, for the theory studied 
in \cite{Marino-Marcello} the corresponding correlator
was found to be
\beq
\langle\mu(x)\mu^{\dag}(y)\rangle=
\frac{e^{-2\pi\rho_{s}(\delta)|x-y|}}{|x-y|^{\alpha(\delta)}},
\label{sk-cf}
\eeq
where the expressions for 
$\rho_{s}(\delta)$ and $\alpha(\delta)$ 
have been carefully determined 
in \cite{Marino-Marcello}. In particular,
$\alpha(\delta)=\left[\frac{64}{\pi^2+16}+
\frac{\alpha_{EM}}{4\pi^{2}}\right]
(n \delta)^2$ with $n = 1$ for 
YBCO and $n = 4$ for LSCO, the 
factor of four being a 
consequence of the existence of 
four branches in the Fermi surface for 
this compound, as discussed in 
\cite{Marino-Marcello}. $\alpha_{EM}$ is the
electromagnetic fine structure constant and we
see that the contribution from the electromagnetic
coupling is negligible. 
The $\rho_{s}(\delta)$ function is given by 
$\rho_{s}(\delta)=\rho_{s}(0)[1 - A\delta^2]$, for YBCO and 
$\rho_{s}(\delta)=\rho_{s}(0)[1 - B\delta - C\delta^2]^{1/2}$, 
for LSCO, and again the 
different behavior being ascribed to 
the form of the Fermi surface in each 
case \cite{Marino-Marcello}. The 
constants $A$,$B$ and $C$ have been 
evaluated from first principles in 
\cite{Marino-Marcello}. In order to obtain the 
$\delta$-dependence of the spin 
stiffness $\rho_{s}$ and of the spinon 
coupling $q$ 
in our model (\ref{Effective-Action}), 
we now match the two correlation functions 
in (\ref{sk-cf1}) and (\ref{sk-cf}) 
(ordered phase), obtaining
$\rho_s=\rho_{s}(\delta)$ and 
$q=[\frac{128}{\pi^2+16}]^{1/2}(n\delta)$. 
The sublattice magnetization in the ordered 
phase is given by $M(\delta)=\sqrt{\rho(\delta)}$, and 
consequently $\delta_{AF}$ can be obtained 
from $\rho(\delta_{AF})=0$, both in good
agreement with experiment, see \cite{Marino-Marcello}. 
For $\delta >\delta_{AF}$, on 
the other hand, 
where $\rho_s = 0$, we shall assume that 
the expression for $q(\delta)$
still holds. This is quite reasonable 
since $q$ was 
introduced by a local canonical 
transformation, and at least locally there 
is short range AF order.

{\it Cooper pair formation.} 
Let us now investigate the conditions for Cooper pairing. 
We shall first introduce in the partition function 
(\ref{Part-Func}) the skyrmion current, 
${\cal J}^{\mu} = \frac{1}{2\pi}\epsilon^{\mu\alpha\beta}
\partial_{\alpha}{\cal A}_{\beta}$, through the identity
\beq
1=\int{\cal D}{\cal J}_{\mu}{\;}
\delta[{\cal J}_{\mu}-\frac{1}{2\pi}\epsilon^{\mu\alpha\beta}
\partial_{\alpha}{\cal A}_{\beta}].
\label{zja}
\eeq
Integrating over $z_i^{\dag},z_i$ and 
$\overline{\psi}_a,\psi_a$, we obtain, 
at leading order, the effective 
Lagrangian
\beq
{\cal L}_{eff}[{\cal A}_{\mu}]=
\frac{N}{2}{\cal A}_{\mu}({\bf x},\tau)
\Pi^{\mu\nu}({\bf x}-{\bf y},\tau-\tau^{\prime})
{\cal A}_{\nu}({\bf y},\tau^{\prime}),
\eeq
where $\Pi^{\mu\nu}({\bf x}-{\bf y},
\tau-\tau^{\prime})$ has Fourier transform
given by $\Pi^{\mu\nu}({\bf p},\rmi\epsilon_{m})=
\Pi^{\mu\nu}_{B}({\bf p},\rmi\epsilon_{m})+
\Pi^{\mu\nu}_{F}({\bf p},\rmi\epsilon_{m})$,
which are respectively the contributions to the finite 
temperature vacuum polarization 
coming from the complex scalar fields $z_i$ 
(Schwinger bosons) and fermions 
$\psi_{\alpha,\lambda}$ (spinons). 

In order to obtain the effective current-current 
interaction between skyrmions, we use an exponential 
representation for the $\delta$-function in 
(\ref{zja}) and integrate over ${\cal A}_\mu$ and 
the corresponding Lagrange multiplier field. 
The result is
\beq
{\cal Z}=\int{\cal D}{\cal J}_{\mu}{\;}
e^{\left\{-2\pi^{2}
\int\rmd^{3}{x}\int\rmd^{3}{y}
{\cal J}_{\mu}({x})
{\Sigma}^{\mu\nu}({x}-{y}){\cal J}_{\nu}({y})\right\}},
\label{Current-Current}
\eeq
where $\Sigma^{\mu\nu}({p})=\Pi^{\mu\nu}({p})/ p^{2}$,
$x=(\tau,{\bf x})$ and $p=(\rmi\epsilon_{m},{\bf p})$.
The real time effective interaction energy between static 
skyrmions $(\epsilon_{m}=0)$ is then 
\beq
{\cal H}_{I}=2\pi^{2}
\int\rmd^{2}{\bf x}\int\rmd^{2}{\bf y}{\;}
\rho({\bf x}){\;}\Sigma^{00}({\bf x}-{\bf y};0){\;}\rho({\bf y}),
\eeq
where $\rho({\bf x}) = {\cal J}_{0}({\bf x})$ 
is the dopant charge density and 
$\Sigma^{00}({\bf x}-{\bf y};0)$ has Fourier 
transform given by 
$\Sigma^{00}({\bf p})=\Pi_{B}({\bf p})+
\Pi_{F}({\bf p})$ with
\bea
\Pi_{B}({\bf p})&=&-\frac{\Delta}{2\pi}
+\frac{1}{2\pi}
\int_{0}^{1}\rmd x{\;}
\sqrt{|{\bf p}|^{2}x(1-x)+m^{2}}  \nonumber \\
&\times& \coth{\left(\frac{\sqrt{|{\bf p}|^{2}x(1-x)+m^{2}}}
{2k_{B}T}\right)},
\label{pib}
\eea
and
\beq
\Pi_{F}({\bf p})=\frac{q^{2}}{\pi}
\int_{0}^{1}\rmd x{\;}\sqrt{|{\bf p}|^{2}x(1-x)}
\tanh{\left(\frac{\sqrt{|{\bf p}|^{2}x(1-x)}}
{2k_{B}T}\right)}.
\label{pif}
\eeq
In the above expressions, $m$ is 
the inverse correlation length of the 
quantum disordered phase of the CP$^{N-1}$ model,
where $\Delta=8\pi\left(1/g_{c}-1/g_{0}\right)$ 
and $\rho_{s}=0$. At order $N$, it 
is given exactly by $m(T)=\Delta+2k_{B} 
T e^{-\Delta/k_{B}T}$ \cite{Starykh}.

For two charges at positions ${\bf x}_{1}$ 
and ${\bf x}_{2}$, we have 
$\rho({\bf x})=\delta^{(2)}({\bf x}-
{\bf x}_{1})+\delta^{(2)}({\bf x}-{\bf x}_{2})$. 
After discarding self-interactions, we 
obtain (${\bf r}={\bf x}_{1}-{\bf x}_{2}$)
\beq
V({\bf r})=\int\rmd^{2}{\bf p}{\;}
\Sigma^{00}({\bf p}){\;}e^{\rmi{\bf p}\cdot{\bf r}}+
V_{l}({\bf r}),
\label{Int-Potential}
\eeq
where we have also introduced the 
centrifugal barrier potential between the two 
charges that form the Cooper pair, 
$V_{l}({\bf r})=l(l+1)\hbar^2/2M^{*}{\bf r}^{2}$, 
with $l$ specifying the relative orbital 
angular momentum of the pair and $M^{*}$
the effective mass of the charges.

{\it Zero temperature limit.}
It is well known that in high-T$_c$ 
cuprates, Cooper pairs form at relatively 
short distances. In this limit, large $|{\bf p}|$, 
we have, at $T=0$
\beq
V({\bf r})\rightarrow \int\rmd^{2}{\bf p}
\left[\frac{1}{8|{\bf p}|} - \frac{2q^{2}}{8|{\bf p}|}\right]
{\;}e^{\rmi{\bf p}\cdot{\bf r}}+V_{l}({\bf r}).
\label{Effective-Potential}
\eeq
The above expression clearly shows 
a competition between the spin fluctuations of 
the Cu$^{++}$ spins (first term) 
and of the O$^{--}$ doped spins (second term). 
For small enough doping, $q^{2}<1/2$, 
the potential is always repulsive and 
there is no charge pairing. For 
$q^{2}>1/2$, on the other hand, the potential has 
a minimum and charge (skyrmion) 
pairing occurs. We conclude that the 
critical doping for the onset of 
superconductivity is determined by the condition 
$q^{2}(\delta_{SC})=1/2$. We observe 
that without the Cu$^{++}$ background, the 
interaction potential (\ref{Effective-Potential}) 
would always have bound states for any $q\neq 0$, 
at zero temperature, and $\delta_{SC}=0$. This is 
what happens in the mean field phase 
diagram of Kotliar and Liu \cite{Kotliar-Liu}. 
We see that the effect of the Cu$^{++}$ background 
is to shift the value of $\delta_{SC}$ to 
its correct position in the phase diagram.

{\it Determination of $\delta_{SC}$.}
From the expression of $q$ in terms of 
$\delta$ ($q = \sqrt{2\alpha(\delta)}$), 
we see that $\delta_{SC}$ is only 
determined by the shape 
of the Fermi surface of the compound. 
In particular, we see that
$\delta_{SC}^{YBCO}=4\delta_{SC}^{LSCO}$, 
a result that is 
verified by experiments, if we 
take in account the relation between $\delta$ and 
the stoichiometric doping parameter 
$x$, namely $\delta=x$ for LSCO and 
$\delta=x-0.20$ for YBCO. Another 
prediction of our model is that compounds 
with similar Fermi surfaces should 
have the same superconducting critical 
doping $\delta_{SC}$. We get 
$\delta_{SC}^{YBCO}=0.318$ and 
$\delta_{SC}^{LSCO}=0.079$, which 
have a fairly good agreement with experiment. 

{\it Disorder.} Disorder may be 
modelled in the ordered N\'eel phase of 
a doped antiferromagnet by 
considering a continuous random distribution of spin
stiffnesses \cite{em}. The effect 
of introducing a Gaussian$\times |\rho - \rho_s|^{\nu -1}$ 
distribution, with exponentially 
suppressed magnetic dilution, in the original 
model \cite{Marino-Marcello}, is 
a correction for $\alpha(\delta)$, namely 
$\alpha(\delta)\rightarrow 
\alpha'(\delta) = \alpha(\delta) + \nu$ \cite{em}. 
Choosing $\nu = \frac{1}{8}$ 
{\it for both compounds}, we get 
$\delta_{SC} = \frac{1}{n} 
\sqrt{ \frac{\pi^2 +16}{512}}$, or equivalently
$x_{SC}^{YBCO} = 0.425$ and 
$x_{SC}^{LSCO} = 0.056$, in good agreement
with experiment. 

{\it Finite temperature.}
For $\Pi_{B}$ we shall expand in $k_{B}T/m$, 
since it is clear that $m(T)>k_{B}T,\forall T$. 
$\Pi_{B}$ will then be simply given by its 
zero temperature limit, where $m=\Delta$. For 
$\Pi_{F}$, on the other hand, such a 
low $T$ expansion is not necessarily valid 
even for $|{\bf p}|\gg k_{B}T$. We will then 
have to split the integral over the Feynman 
parameter $x$ in (\ref{pif}) into three parts. For 
$0\leq x \leq x_{c}$ and $1-x_{c}\leq x \leq 1$, 
$x_{c}=(k_{B}T/|{\bf p}|)^{2}$, we will 
use a high $T$ expansion, while for 
$x_{c} \leq x \leq 1-x_{c}$ we use 
the low $T$ expression. We obtain
\begin{equation}
\Sigma^{00}({\bf p})=\frac{(1-2q^{2})}{8 |{\bf p}|}-
\frac{m}{\pi|{\bf p}|^{2}}+
\frac{m^{2}}{2|{\bf p}|^{3}}+
\frac{16 q^{2} T^{3}-4 m^{3}}{3\pi|{\bf p}|^{4}}.
\eeq
Inserting this in (\ref{Int-Potential}), we get
$V({\bf r})$, and from the treshold conditions 
for the formation of bound states, namely 
$V^{\prime}({\bf r}_{0})=0$ and 
$V^{\prime\prime}({\bf r}_{0})=0$, we obtain
the relation
\bea
(k_{B}T_{SC})^{3}&=&-\frac{\pi(1-2q^{2})
\alpha^{3}}{512 q^{2}}+\frac{m\alpha^{2}}{32 q^{2}}-
\frac{3\pi m^{2}\alpha}{128 q^{2}}+\frac{m^{3}}{4 q^{2}} \nonumber \\
&-&\frac{3\pi^{2}l(l+1)\alpha^{4}}{q^{2}M^{*}v_{F}^{2}},
\label{SC-Line}
\eea
where $\alpha=\hbar v_{F}/{\bf r}_{0}$, $m = \Delta$, 
and ${\bf r}_{0}$ is the minimum of the potential
(it also measures the size of the Cooper pair). 

%
\begin{figure}[ht]
\centerline{\epsfxsize=8cm \epsffile{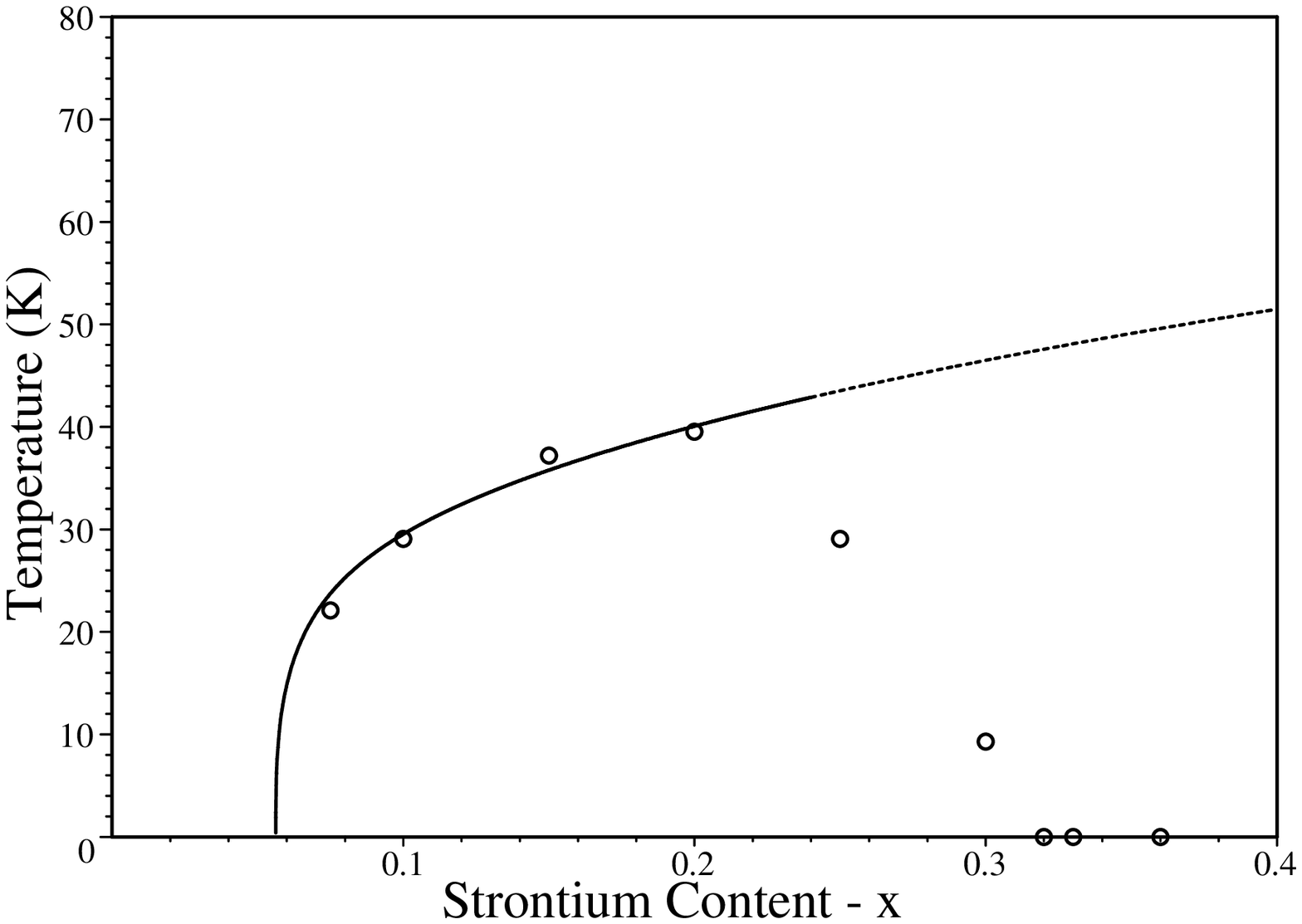}}
\vskip0.2cm
\caption{Plot of curve (\ref{SC-Line}) for {\lsco}. 
Experimental data from \protect\cite{Keimer}.}
\label{fig1}
\end{figure}
%
%
\begin{figure}[ht]
\centerline{\epsfxsize=8cm \epsffile{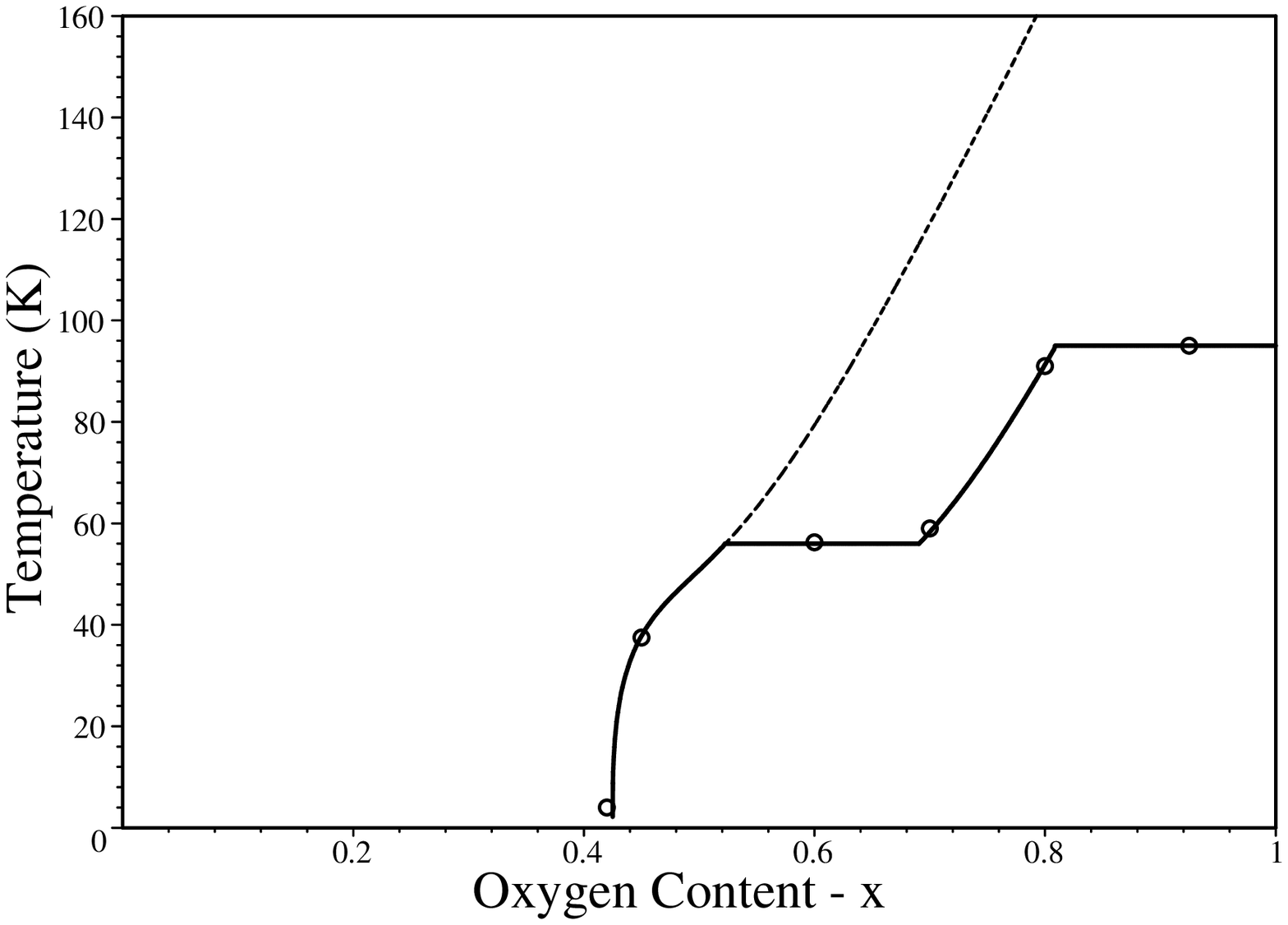}}
\vskip0.2cm
\caption{Plot of curve (\ref{SC-Line}) for {\ybco}. 
Experimental data from \protect\cite{Rossat-Mignod}.}
\label{fig2}
\end{figure}
%

{\it Comparison with experiment.}
In order to make contact with experimental data we need the 
doping dependence of $\Delta$. For YBCO, we use
$\Delta(\delta)=\Delta_{0}[(\delta/\delta_{AF})^{2}-1]$,
in agreement with the results of \cite{Marino-Marcello},
with $\Delta_{0}=8.8$ meV. For LSCO, we shall use an expression 
that fits the experimental data of \cite{Keimer}, namely 
$\Delta(\delta)=\Delta_{0}[(\delta/\delta_{AF})^{2}-1]^{1/2}$, 
with $\Delta_{0}=1.0$ meV. For the $T=0$ AF quantum critical point
$\delta_{AF}$, we know from experiments that $\delta_{AF}=0.22$ 
for YBCO and $\delta_{AF}=0.02$ for LSCO. Inserting in (\ref{SC-Line}) 
the values of $\delta_{SC}$ at $T=0$, obtained previously, 
we get a relation that fixes $M^{*}v_{F}^{2}$ with respect to 
$r_{0}$. In figs. \ref{fig1} and \ref{fig2} we plot the curve 
(\ref{SC-Line}) for LSCO and YBCO, respectively, with $r_{0}=38$ 
{\AA}, $\hbar v_{F}=0.18$ eV {\AA} for LSCO and 
$\hbar v_{F}=1.08$ eV {\AA} for 
YBCO, and $l = 2$ ($d$-vave pairing). In the first plot (LSCO), the 
dashed part is in the region where $T>T^{*}$ and we should move 
to a new saddle-point. In the second plot (YBCO) we have shifted the 
curve (dashed part) to the right in the regions $\delta=[0.52,0.7]$ 
and $\delta=[0.8,1]$ in order to comply with the effects of the 
out-of-plane {\oco} chains, which produce the observed $60$ K
and $90$ K plateaus, where the extra holes do not enter in the 
{\cuoo} planes. Furthermore, for YBCO, $T^{*}$ is higher than
$T_{max}$ ($\delta = 1$) and therefore imposes no restrictions to our results.      

{\it Final remarks.}
We would like to remark that our 
theory (\ref{Effective-Action}) also gives a 
simple interpretation for the pseudogap phenomena. Indeed,
for $T_{SC}<T<T^{*}$ spinons are paired into $d$-wave
singlets but holons (skyrmions) repel each other and
there is no superconducting state. Only for $T<T_{SC}$
we do have Cooper pair formation and superconductivity. 

\end{narrowtext}


This work has been supported in part by  FAPERJ and 
PRONEX - 66.2002/1998-9. E. C. M. was partially supported by CNPq
and M. B. S. N. by FAPERJ.


\end{document}